\documentclass[conference]{IEEEtran}
\IEEEoverridecommandlockouts
\usepackage{cite}
\usepackage{amsmath,amssymb,amsfonts}
\usepackage{algorithmic}
\usepackage{graphicx}
\usepackage{float}
\usepackage{subfigure}
\usepackage{textcomp}
\usepackage{xcolor}

\def\BibTeX{{\rm B\kern-.05em{\sc i\kern-.025em b}\kern-.08em
    T\kern-.1667em\lower.7ex\hbox{E}\kern-.125emX}}
\begin{document}

\title{BSRA: Block-based Super Resolution Accelerator with Hardware Efficient Pixel Attention}

\author{\IEEEauthorblockN{Dun-Hao Yang, and Tian-Sheuan Chang, \textit{Senior Member, IEEE}}
\IEEEauthorblockA{\textit{Institute of Electronics, National Yang Ming Chiao Tung University,} \\
Hsinchu, Taiwan \\
}
}
\maketitle

\begin{abstract}
Increasingly, convolution neural network (CNN) based super resolution models have been proposed for better reconstruction results, but their large model size and complicated structure inhibit their real-time hardware implementation. Current hardware designs are limited to a plain network and suffer from lower quality and high memory bandwidth requirements. This paper proposes a super resolution hardware accelerator with hardware efficient pixel attention that just needs 25.9K parameters and simple structure but achieves 0.38dB better reconstruction images than the widely used FSRCNN. The accelerator adopts full model block wise convolution for full model layer fusion to reduce external memory access to model input and output only. In addition, CNN and pixel attention are well supported by PE arrays with distributed weights. The final implementation can support full HD image reconstruction at 30 frames per second with TSMC 40nm CMOS process.


\end{abstract}


\section{Introduction}
\begin{figure}[t]
\centering
\includegraphics[width=0.4\textwidth]{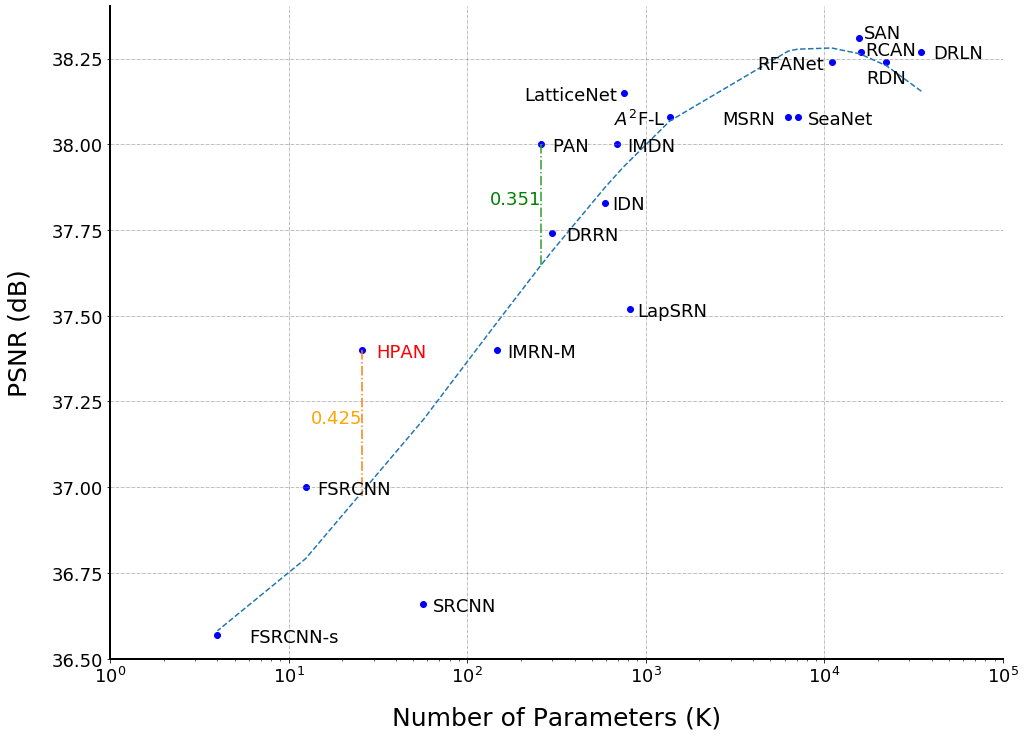}
\caption{Image quality vs model size for different SR models}
\label{performance versus PSNR}
\end{figure}

Convolutional neural network (CNN) based super resolution (SR) \cite{dong2015image} is getting popular in recent years because of better reconstructed high resolution images over traditional interpolation methods. These SR models are getting deeper and wider and use complicated structure to get better performance\cite{zhang2018image, ahn2018fast, dai2019second, zhao2020efficient} as shown in Fig.~\ref{performance versus PSNR}. However, SR models suffer from high computational complexity and memory bandwidth because their feature sizes are not getting smaller over layers and they need to process large input size. Thus, hardware acceleration is demanded for real-time applications.

Various hardware accelerators have been proposed\cite{9223656, Yen2020RealtimeSR, 9159619}. However, current designs only use plain networks like the widely used FSRCNN\cite{dong2016accelerating} instead of recent large size and complicated models due to the high model complexity, which limits their reconstructed image quality. In addition, most of the designs use layer-by-layer processing that will need to store intermediate data to DRAM and load it back for each layer or large buffer size, which is extremely bad for SR applications due to their large feature size.

To address the above problems, we propose a full model block convolution based super resolution accelerator (BSRA) with hardware efficient pixel attention (HPAN) model. The HPAN model uses pixel attention for better image quality than FSRCNN but keeps the structure simple for small model size and low complexity. The accelerator uses block convolution for the whole model instead of certain layers as in \cite{li2021block} to enable the whole model fusion to reduce the external memory bandwidth to model input and output only and needs a small on-chip buffer size. The required convolution and pixel attention are well supported by the proposed distributed weight PE array. The final implementation can achieve real-time full HD image throughput with TSMC 40nm CMOS process.

\section{Proposed SR Model with Hardware Efficient Pixel Attention}

\begin{figure*}[t]
\centering
\includegraphics[scale=0.4]{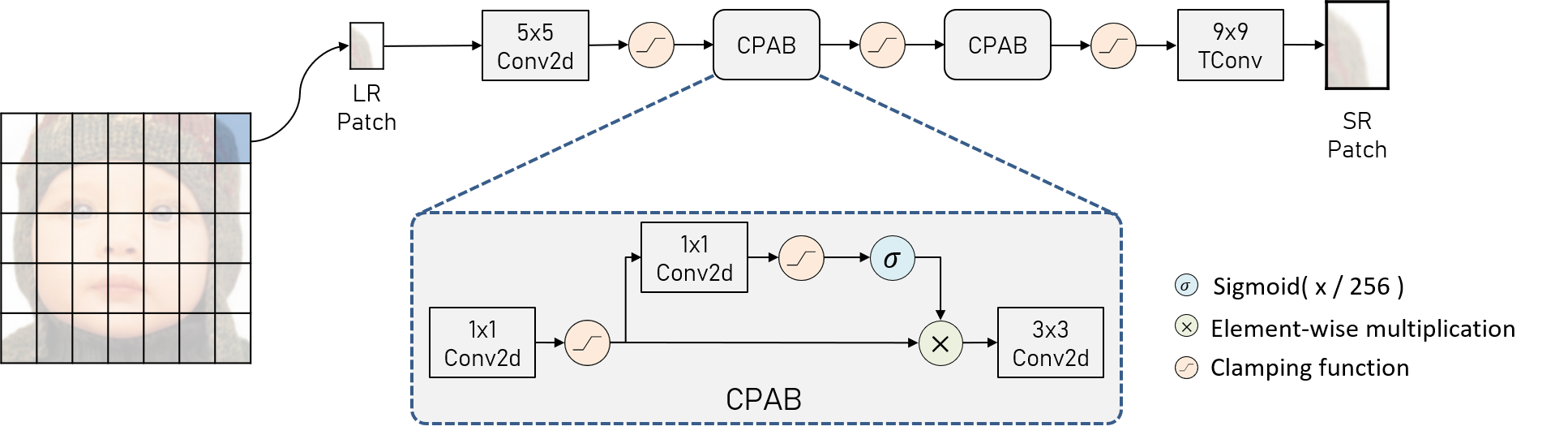}
\caption{The proposed HPAN model}
\label{The block diagram of proposed neural network}
\end{figure*}

\subsection{Related Work}

Since the first CNN based SR model, SRCNN~\cite{dong2015image}, was proposed, many deep learning skills are added to the super resolution model to get better image quality. CARN~\cite{ahn2018fast} cascades the residual in residual blocks to get multilevel representations. RCAN~\cite{zhang2018image}, which is over 400 layers, uses residual blocks and channel attention to learn more channel-wise information. SAN~\cite{dai2019second} proposes non-locally enhanced residual group to gain long distance and structure message. These models are getting deeper and wider for better images.

Fig.~\ref{performance versus PSNR} the scatter chart of the SR model size and its image quality. This figure shows that the quality of the reconstructed image is increasing with model size, but the increment of parameters is not efficient at million scale. Take FSRCNN~\cite{dong2016accelerating} and RCAN~\cite{zhang2018image} for example, the difference in performance on Set5  between them is only 1.3 dB, but the model size of the latter one is about 1300 larger than the former one. The model size increment is very inefficient so that most of hardware designs \cite{9223656, Yen2020RealtimeSR, 9159619} use plain and shallow networks just like SRCNN~\cite{dong2015image} and FSRCNN~\cite{dong2016accelerating}. For example, in ~\cite{9159619}, they uses a classification network to judge whether image tiles are with more high frequency signal and send them to appropriate plain networks. This mechanism can reduce the computation overhead of the SR part, but the classification part is much bigger than the SR part. Therefore, this paper tries to find a better architecture which has a trade-off between model size and reconstructed image quality for single image super resolution.

\subsection{Proposed HPAN Model}
Fig.~\ref{The block diagram of proposed neural network} shows the proposed HPAN model based on convolution pixel attention mechanism and the number of parameters is just about 26K. The model design is based on the following observations. First, for the overall model structure, skip or dense connection is popular in most of the models to learn the residual between high resolution (HR) images and SR images, but this will need extra memory access for hardware implementation. Thus, instead of the residual architecture, we take the attention mechanism into consideration. In Fig.~\ref{performance versus PSNR}, RCAN~\cite{zhang2018image} and SAN~\cite{dai2019second} are the best two models in performance with different attention methods. However, the trend line shows that pixel attention (PAN)~\cite{zhao2020efficient} is better. Thus, we use pixel attention to construct our model. However, ~\cite{zhao2020efficient} has a local skip connection in the basic block and a global interpolated image addition. Besides, in the up-sampling stage, the final image is not generated immediately because the interpolated features still need to be tuned by some convolution layers and thus needs large computation. These are not hardware friendly.

Based on the above observations, we propose the hardware-friendly pixel attention model with 26K parameters as shown in Fig.~\ref{The block diagram of proposed neural network}. The model consists of three stages: feature extraction, mapping, and reconstruction. The feature extraction stage uses one convolution layer. The mapping stage is based on the Clamping Pixel Attention Block (CPAB) in (\ref{eq:cpab}) to extract features. After cascading some CPABs, we use transpose convolution to reconstruct the image directly to prevent the problem mentioned before. 

The representation of CPAB in Fig.~\ref{The block diagram of proposed neural network} is shown as follows.
\begin{equation}
\label{eq:clamping}
    Clamp(f_{in}) =
    \left\{
             \begin{array}{lr}
             +255, & f_{in} \geq +255  \\
             -255, & f_{in} \leq -255 \\
             f_{in}, & \text{otherwise}
             \end{array}
    \right.
\end{equation}

\begin{equation}
\label{eq:sigmoid}
    D(f_{in}) = sigmoid(f_{in}/256)
\end{equation}

\begin{equation}
\label{eq:pa}
    PA(f_{in}) = M_{ew}(f_{in},Clamp( D(CL_{1}(f_{in}))))
\end{equation}

\begin{equation}
\label{eq:cpab}
    CPAB(f_{in}) = CL_{3}( PA( CL_{1}( f_{in} ) ) )
\end{equation}
where $f_{in}$ is input feature, and $Clamp()$, $D()$, $M_{ew}(,)$, $PA()$, and $CL_{i}()$ denote clamping function, divisor, element-wise multiplication, pixel attention and convolution with kernel size $i$, respectively. In this model, with (\ref{eq:sigmoid}), the weight and activation distribution will be more reasonable after training because the range of pixel intensity is from 0 to 1 or 0 to 255. This is applied to every convolution to ensure the proper range. Besides, since the value of the attention map is smaller than 1, the absolute value of multiplication results will not exceed 255. The divisor in (\ref{eq:sigmoid}) chooses 256 for hardware-friendly design as a simple shifting operation.

\section{Proposed Architecture}

\subsection{Overview}

\begin{figure}[t]
\centering
\includegraphics[width=0.45\textwidth]{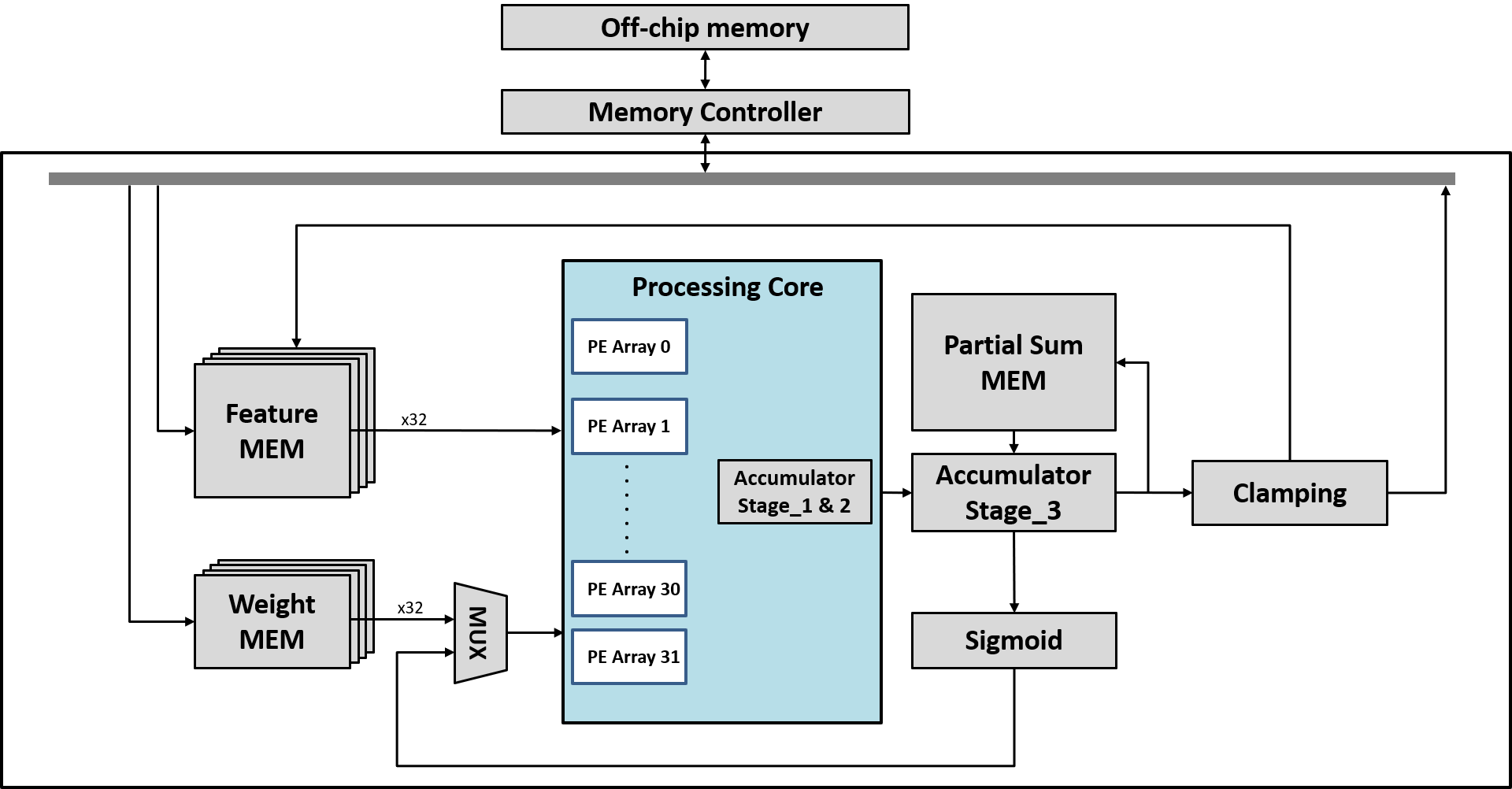}
\caption{System architecture}
\label{System architecture}
\end{figure}

Fig.~\ref{System architecture} shows the proposed system architecture. This design gets weights and input images by accessing external memory and stores them into weight and feature SRAM buffers. Both have 32 banks to store kernels of different layers and intermediate results, respectively, for the computing core. This core consists of 32 processing element (PE) arrays to process 32 channels of input. Then the outputs of different channels are summed in a local accumulator and stored in the partial sum buffer. These partial sums will be further accumulated in the selective adder for convolution. The results will be further processed by sigmoid and clamping function according to the proposed model. The results will be sent to the multiplexer for attention mechanism or feature memory for the next layer.


\begin{figure}[t]
\centering
\includegraphics[width=0.45\textwidth]{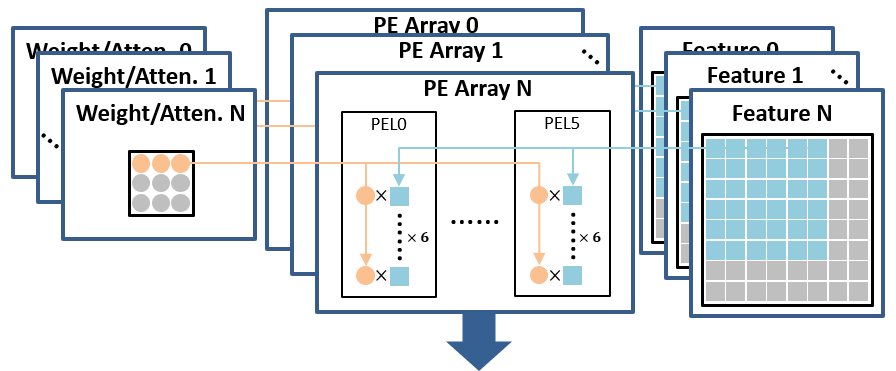}
\caption{PE arrays}
\label{PE array}
\end{figure}

\subsection{PE Array with Distributed Weight}
Fig.~\ref{PE array} shows the proposed one PE array, which consists of six PE lines (PEL0 to PEL5) with six PEs in each PELx. Each PE is a multiplier with registers. In the figure, each PE array is for one channel of convolution in a layer. In this way, the input feature maps and weights share the same spatial location in the same PE array. Therefore, the multipliers in each PE get different features according to their location in the spatial space. In other words, 36 input pixels are allocated to 36 PEs. As for the multiplicands in one PE array, there are two choices, one is convolutional kernel weights and the other is pixel-wise attention mask. If the multiplicands are kernel weights, the weights will broadcast vertically. If the multiplicands are an attention mask, the mask values will be distributed in PE array. With this distributed weight scheme, this design can support convolution as well as pixel attention. After completing the multiplications, the 36 results are transferred to the next stage to be accumulated.


\begin{figure}[t]
\centering
\includegraphics[width=0.4\textwidth]{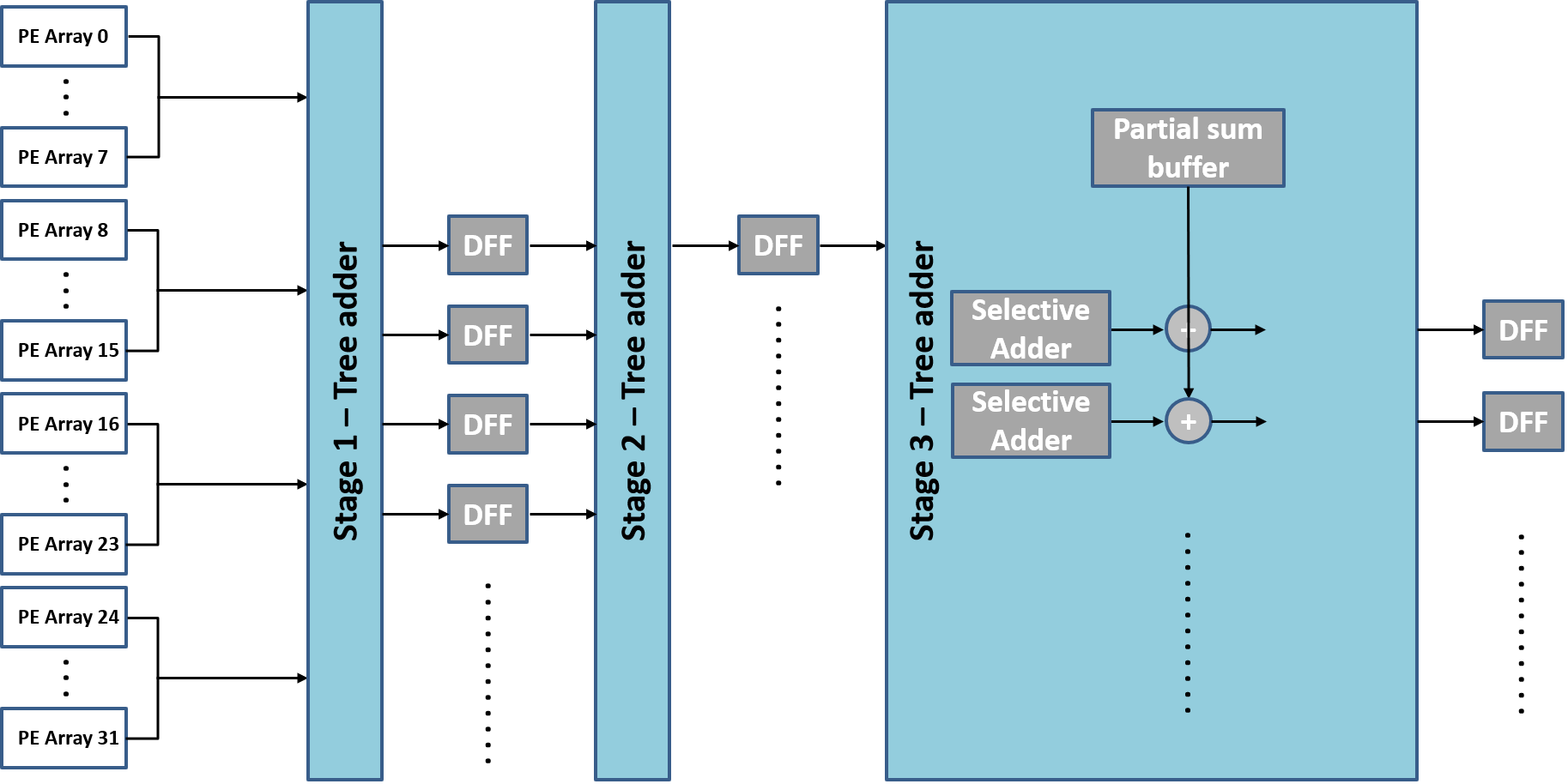}
\caption{The three-stage accumulator.}
\label{Accumulator_stage}
\end{figure}

\subsection{Three-Stage Accumulator}
Fig.~\ref{Accumulator_stage} shows the accumulator to accumulate the partial sum, which is partitioned into three-stage pipelines for minimizing the critical path and attention mechanism. In which, for our 32 channel computation, eight partial sums at the same spatial location but different channels are summed up at the first stage, and their results are further accumulated to complete a 32 channel summation at the second stage. More channel summations are done at the third stage for the final convolutional results. The selective adder at the third stage will select the proper input based on convolution or attention operation for the correct output.

\begin{figure}[t]
\centering

\includegraphics[width=0.48\textwidth]{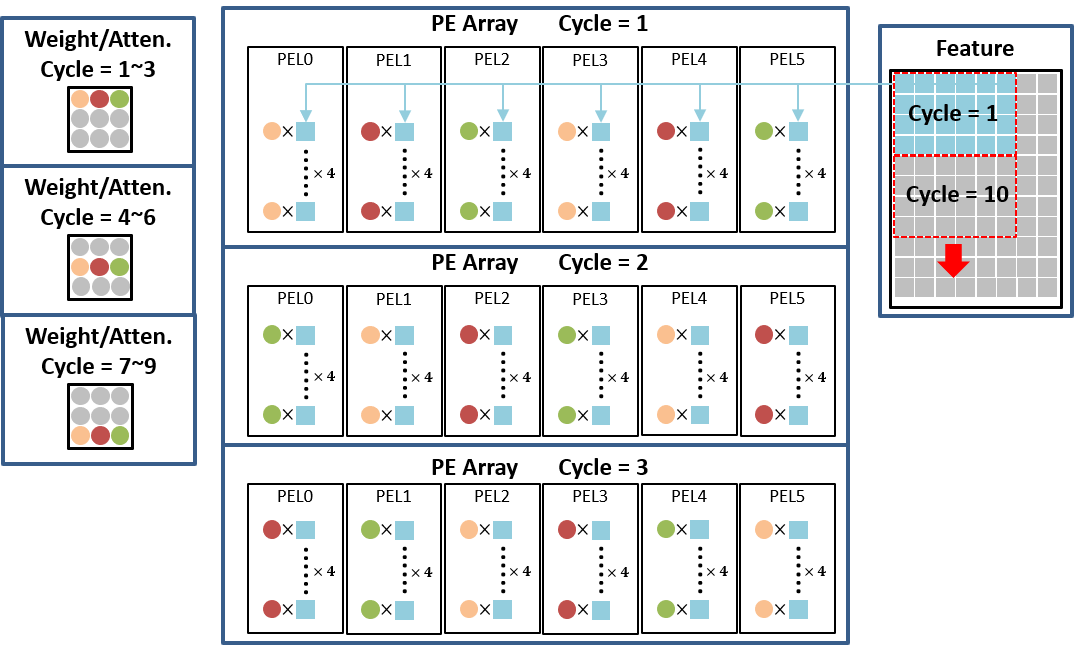}

\caption {An example of the proposed data flow. In which, for data allocation, the same colored geometric shapes share the same weights. If kernel size is $k$, after $k$ clock cycle, the cached feature should be updated.}
\label{Data flow chart}
\end{figure}


\begin{figure*}[t]
\centering
\includegraphics[width=0.8\textwidth]{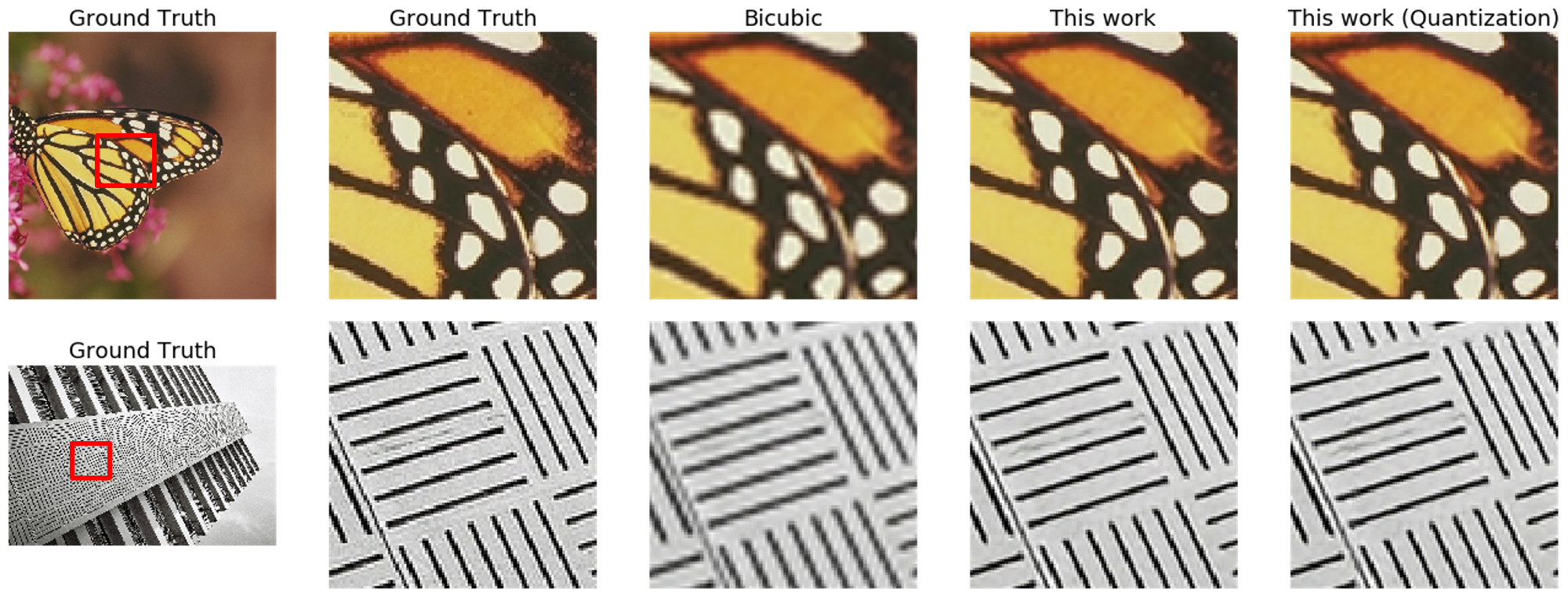}
\caption{Comparison of reconstructed images from the Set14 and Urban100 datasets. The images from left to right are original ground truth, enlarged ground truth, bicubic interpolation, our floating-point result and quantization result with block convolution, respectively.}
\label{image results}
\end{figure*}

\subsection{Data Flow of Convolution}
Fig.~\ref{Data flow chart} shows an example data flow with one PE array with 4 PEs in each PEL for simplicity. In the proposed data flow, the kernel weights are broadcast vertically along the PE array row and the feature inputs are transferred to PEs depending on their spatial locations. With this allocation, the results of PEs, $P^{sum}_{i,j}$, are belong to the output pixel $O_{i,j}$ at the first cycle. At the next cycle, the weights in the PE array are circularly shifted right. The results $P^{sum}_{i,j}$ are belong to $O_{i-1,j}$. At the third cycle, the weights are shifted again and the results are belonging to $O_{i-2,j}$. When the first row of weights are fully used, the second row of weights are loaded into the PE array. By repeating this processing flow, all partial sums can be generated. These partial sums will be accumulated in the accumulator and stored in the partial sum buffer. After nine clock cycles, the feature input will be shifted down by four pixels and will be loaded for another round of computation until the bottom of this tile. After that, the feature input will be shifted right by six pixels for computation. With this, the feature map can be reused to reduce the memory bandwidth due to the large input in the SR model. 


\subsection{Block Convolution for Whole Model Fusion Execution}
Because the feature size of intermediate layers is as large as the input of the SR model, the required on-chip buffer size will grow up quickly as the input size is increased. This problem cannot be solved well by commonly used tile-based processing \cite{9159619} due to large boundary data storage or computation between tiles. Thus, this paper adopts the block convolution  ~\cite{li2021block} by splitting the input into nonoverlapped tiles and processing them with suitable padding without boundary information. However, the previous approach in \cite{li2021block} limits this to only certain layers and needs a large tile size due to information loss at the boundary. In contrast, the proposed network with pixel level attention could reduce this impact to minimum. We apply this to all layers with small tile size ($40,48$) , and enable the whole model fused execution by processing one tile from input to output directly. All intermediate data can be stored in small on-chip buffers without external DRAM access as other SR accelerators. With this, the external DRAM access can be reduced to the model input and output only.


\section{Experimental Results}

\subsection{Model Simulation Result}
Fig.\ref{image results} shows the simulation results and comparisons. The results with bicubic interpolation will lose a lot of detail and become blur, especially at the stripe boundary. In contrast, the proposed network can reconstruct images correctly and the generated results are more clear than the traditional ones. The results are almost kept the same even when we quantize the weights and activations and include the full model block convolution without handling boundary information.

Table~\ref{comparison with others} shows performance comparisons with other works. In which, VDSR and PAN have better performance than ours, but they also need 10 and 25 times of model parameters than ours. The performance difference between VDSR and ours is marginal for several datasets. In contrast, our model size is close to the lightweight models but has much better performance. Our model has achieved a good balance between model size and performance, which has also been illustrated in Fig.~\ref{performance versus PSNR}.



\begin{table}[t]
\caption{Comparison of several software approaches on commonly used datasets (Urb.100 is Urban100 and M.109 is Manga109).}
\centering
\begin{tabular}{lllllll}
\hline
Algorithm & Param. & Set5                              & Set14                             & B100                              & Urb.100                          & M.109                           \\
\hline
Bicubic   &   -        & 33.66                             & 30.25                             & 29.57                             & 26.89                             & 30.86                              \\
SRCNN-Ex  & 57,184      & 36.66                             & 32.45                             & 31.36                             & 29.50                             & 35.60                              \\
FSRCNN    & 12,464      & 37.00                             & 32.63                             & 31.50                             & 29.88                             & 36.67                              \\
This work & \textbf{25,920}      & \textbf{37.38} & \textbf{32.91} & \textbf{31.69} & \textbf{30.29} & \textbf{37.00}  
\\ 
\hline
VDSR      & 665,000     & 37.53           & 33.05           & 31.90           & 30.77           & 37.22            \\
PAN       & 261,000     & 38.00            & 33.59            & 32.18            & 32.01            & 38.70             \\

\hline
\end{tabular}
\label{comparison with others}
\end{table}

\begin{table}
\centering
\caption{Comparison with other hardware works}
\begin{tabular}{|c|c|c|c|} 
\hline
Work                & \cite{Yen2020RealtimeSR}        & \cite{9159619}                                                                      & This work        \\ 
\hline
Method              & Improved IDN & S/L FSRCNN                                                                 & Pixel Attention  \\ 
\hline
Resolution   & FHD          & FHD                                                                        & FHD              \\ 
\hline
Process & 32 nm        & 65 nm                                                                      & 40nm             \\ 
\hline
Frequency & 200 MHz & 200 MHz & 471 MHz \\
\hline
Parameters          & 12,726       & 16,660\footnotemark[1] & 25,920           \\ 
\hline
PSNR (dB)           & 36.96        & 37.06                                                                      & 37.18            \\ 
\hline
Precision(W/A)      & - / 12   & 8 / 16                                                             & 11 / 18  \\ 
\hline
Memory Size    & -            & 572 KB                                                                       & 232 KB              \\ 
\hline
Gate Count          & 2614K        & -                                                                          & 2837K            \\ 
\hline
External BW & I/O +  & I/O + partial  & I/O only \\
            &intermediate & intermediate & \\
 \hline

FPS                 & 60           & 31.8                                                                       & 30               \\
\hline
\end{tabular}
\label{Comparison with other hardware works}
\footnotemark[1]{SR model only without classification model}\\

\end{table}

\subsection{Hardware Implementation}

Table~\ref{Comparison with other hardware works} shows hardware implementation results of TSMC 40nm CMOS process and comparison. This design can achieve real-time full HD image reconstruction with 30 frames per second when running at 471 MHz. Compared to other designs, this design can save significant intermediate feature map I/O to DRAM with smaller on-chip buffer size and better reconstruction image quality.

\section{Conclusion}
This paper proposes an SR accelerator with hardware efficient pixel attention model to reconstruct images with 37.38 dB PSNR on Set5 using 25.9 K parameters, which is the first hardware work without just plain network. Besides, we also reduce the external memory access to model I/O only with full model block convolution and layer fusion. Such convolution and pixel attention computation have been well supported by PE arrays with distributed weights. The final implementation can achieve real-time full HD image reconstruction with smaller buffer size and bandwidth but better image quality than other SR accelerators.


\section*{Acknowledgment}
This work was supported by the Ministry of Science and Technology, Taiwan, under Grant 109-2634-F-009-022, 109-2639-E-009-001, 110-2221-E-A49-148-MY3 and 110-2622-8-009-018-SB, and TSMC.

\bibliographystyle{IEEEtran}
\bibliography{citation.bib}
\end{document}